\documentclass{elsart}
\usepackage{amssymb}
\usepackage{graphicx}
\usepackage{cite}
\usepackage{subeqn}
\begin{document}
\begin{frontmatter}
%
\title{Recoiling DNA Molecule: Simulation \& Experiment}
%
\author{Jos\'e Coelho~Neto\corauthref{ca}},
\corauth[ca]{Corresponding author.}
\ead{rabbit@fisica.ufmg.br}
\author{Ronald Dickman},
\ead{dickman@fisica.ufmg.br}
\author{O. N. Mesquita}
\ead{mesquita@fisica.ufmg.br}
\address{\small Departamento de F\'\i sica, ICEx, Universidade Federal de Minas Gerais\\ Caixa Postal 702, CEP 30123-970, Belo Horizonte, Minas Gerais, Brasil}

\date{1 April 2004}

\begin{abstract}
%
Single molecule DNA experiments often generate data from force versus extension measurements involving the tethering of a microsphere to one end of a single DNA molecule while the other is attached to a substrate. We show that the persistence length of single DNA molecules can also be measured based on the recoil dynamics of these DNA-microsphere complexes if appropriate corrections are made to the friction coefficient of the microsphere in the vicinity of the substrate. Comparison between computer simulated recoil curves, generated from the corresponding Langevin equation, and experimental recoils is used to assure the validity of data analysis.
\end{abstract}
\begin{keyword}
DNA \sep single molecule \sep persistence length \sep recoil dynamics
%
\PACS 87.14.Gg \sep 87.15.La \sep 87.80.Cc \sep 05.10.Gg
\end{keyword}
\end{frontmatter}
%
%
\section{Introduction}
\label{intro}

One of the most important characteristics of the DNA molecule is its high mechanical flexibility. Thanks to this, the entire genome of each living organism, which can reach several meters in length, fits inside each of its cells. The flexibility of the DNA molecule also plays a key role in all its cellular functions, such as replication, transcription and recombination. With advances in micromanipulation techniques during the last decade single molecule DNA experiments have become common, and are explored through many different approaches \cite{Strick1}.

Once free in solution, single DNA molecules present entropic elastic behavior, well described, in both low and high stretching regimes, by the \emph{worm-like chain} (WLC) model \cite{Bustamante1, Marko1, Bouchiat1}. The flexibility of the polymeric chain can be inferred through the ratio \(A/L\), where \(A\) is its \emph{persistence length} and \(L\) its \emph{contour length}. Smaller \(A/L\) ratios correspond to more flexible chains. For the DNA molecule, a complete chain can have contour length L varying from around \(2\mu m\) to more than \(1m\), while the persistence length \(A\) is of the order of \(40-50nm\) or \(120-150\) base pairs.

\begin{figure*}
\centering
\includegraphics[width=13cm]{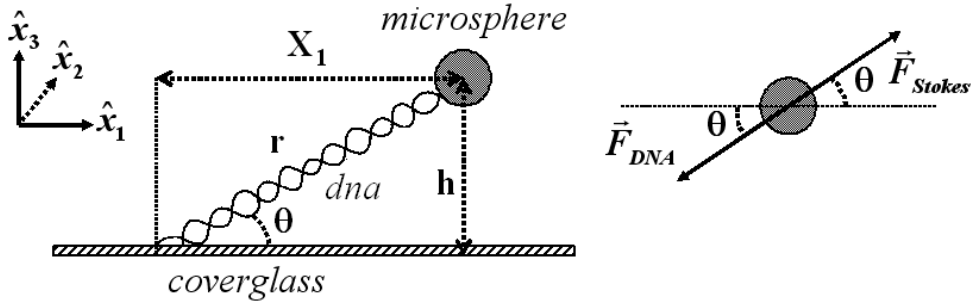}
\caption{Microsphere anchored to a coverglass surface by a single tethered DNA molecule.}
\label{sample}
\end{figure*}

The usual method to obtain \(A\) employing an optical trap \cite{Ashkin1} is based on \emph{force versus extension} measurements \cite{Wang1, Shivashankar1, Shivashankar2, Viana1}. One end of a DNA molecule is attached to a coverglass surface while the other end is tethered to a polystyrene microsphere \emph{(Fig.~\ref{sample})}. The optical trap can then be used to capture and move the microsphere, stretching the DNA molecule. Obtaining force versus extension data for a DNA molecule, however, requires a very precise characterization and calibration of the trap, allowing the determination of the trapping force as a function of the position of the trapped microsphere. Here we examine an alternate method to determine the persistence length \(A\): once the DNA is stretched, one removes the trap and follows the motion of the microsphere in time, using videomicroscopy \cite{Shivashankar2}.

Using this method, Feingold \cite{Feingold1} obtained a value of \(152nm\) for the persistence length of DNA, approximately three times larger than the value obtained from force versus extension measurements. The proposed explanation for this discrepancy was that during the initial part of the recoil, when the DNA is more than \(80\%\) stretched, the motion cannot be considered quasistatic, such that the force at each position is time dependent. Recently, Bohbot-Raviv \emph{et al} \cite{Bohbot-Raviv1} considered a complete nonequilibrium theory for the relaxation of highly stretched semiflexible polymers like DNA and found a better agreement with Feingold's experiment.

We have performed recoil experiments in which the friction between the microsphere and the coverglass is carefully taken into account and found persistence lengths comparable to the values obtained from force versus extension measurements just using the standard quasistatic approximation, without the need of a nonequilibrium theory for the relaxation of the DNA. We also perform computer simulations of the recoil using the corresponding Langevin equation to assure the validity of the assumptions used in the data analysis.

The balance of this paper is organized as follows. In section~\ref{exptech} we describe the recoil method and the motion of the microsphere. Section~\ref{sim} details the simulation of the recoil dynamics of the microsphere. The procedures adopted to obtain experimental recoil curves are described in section~\ref{exp}. In section~\ref{analysis} we describe the analysis of simulated and experimental recoils leading to our results. Conclusions and final remarks are presented in section~\ref{conclusion}.

\section{Experimental Technique}
\label{exptech}

In essence the recoil technique introduced by Shivashankar \emph{et al} \cite{Shivashankar2} is very simple. As in the force-based experiments, optical tweezers are used to find and trap a microsphere anchored to a coverglass surface through a single tethered DNA molecule. The trapped microsphere is moved, stretching the DNA molecule. The tweezers are then cut off suddenly, releasing the microsphere, which is dragged through the fluid by the recoiling DNA molecule. Following the recoil of the microsphere using videomicroscopy, we obtain a recoil curve \(R(t)\), from which we can extract information about \(A\) and \(L\).

According to \cite{Shivashankar2}, the recoil movement can be considered \emph{quasistatic} since the equilibration time for the DNA molecule is very short compared with the time scale for the motion of the microsphere. This means that, during relaxation, the DNA essentially passes through a sequence of equilibrium stretched states \cite{Perkins1}. Because the time resolution of our videomicroscopy experiments (\((1/30) s\)) is large compared to the ratio \(m/\gamma\) (\(\approx 10^{-7} s\)) for the microsphere, inertial effects are negligible in the analysis of the recoils. As a result, the average motion of the microsphere can be described by
\begin{equation}
\gamma\frac{d\left<R\right>}{dt} = -F_{dna}(r),
\label{eqm}
\end{equation}
where \(\left<R\right>\) is the average position of the microsphere, \(r\) is the distance between the extremities of the DNA molecule, \(\gamma\) is the Stokes friction coefficient and \(F_{dna}(r)\) is the entropic elastic force for the DNA molecule given by the WLC model \cite{Bustamante1, Marko1, Bouchiat1},
\begin{equation}
F_{dna}(r) = \frac{k_bT}{A}\left[\frac{r}{L} + \frac{1}{4\left(1 -
\frac{\displaystyle r}{\displaystyle L}\right)^2} - \frac{1}{4} + \sum_{i=2}^{i\leq 7}\alpha_i\left(\frac{r}{L}\right)^i\right],
\label{fdna}
\end{equation}
where the final term represents the correction introduced by Bouchiat \emph{et al} in \cite{Bouchiat1} with \(\alpha_2 = -0.5164228\), \(\alpha_3 = -2.737418\), \(\alpha_4 = 16.07497\), \(\alpha_5 = -38.87607\), \(\alpha_6 = 39.49944\) and \(\alpha_7 = -14.17718\). The bouyancy force on the microsphere is \(5fN\), negligible compared to \(F_{dna}\), which is of the order of \(pN\) during most of the recoil movement.

Far from any boundary, the Stokes friction coefficient between the microsphere and the surrounding medium is
\begin{equation}
\gamma = 6\pi\eta a,
\label{gamma}
\end{equation}
where \(\eta\) is the viscosity of the medium and \(a\) is the radius of the microsphere. There are, however, important considerations to be made regarding the use of Eq.~(\ref{eqm}). The wall-drag effect caused by the coverglass nearby changes the friction coefficient \(\gamma\) of the microsphere \cite{Feitosa1}. As a result, \(\gamma\) can no longer be considered constant and isotropic, and must continuously be corrected as the microsphere moves. This correction depends on the height \(h\) of the microsphere with respect to the coverglass below.

We can integrate Eq.~(\ref{eqm}) directly. However, we cannot obtain a simple explicit analytic expression for \(R(t)\), forcing us to devise an alternative procedure to fit the experimental data. We chose to apply a discrete derivative to the experimental recoil curves, generating data that may be fit directly with Eq.~(\ref{eqm}). Since the microsphere undergoes Brownian motion while recoiling, the numerical derivative will, however, be noisy. We therefore smooth the recoil curve, minimizing the Brownian noise, prior to performing the numerical derivative. To test the validity and effects of the adopted procedure, we compare our result with the simulated the recoil dynamics of a tethered microsphere.

\section{Simulation}
\label{sim}

To simulate the recoil dynamics of the microsphere we solve, numerically, the Langevin equation for the system, using \(F_{dna}\) as an external force \cite{Kampen1, Doi1}. In the non-inertial limit, we have
\begin{equation}
\frac{dR}{dt} = -\frac{F_{dna}(r)}{\gamma} + \frac{\mathcal{F}(t)}{\gamma},
\label{leni}
\end{equation}
where \(\mathcal{F}(t)\) is the Brownian force, which has the properties \(\left<\mathcal{F}(t)\right>=0\) and
\(\left<\mathcal{F}(t)\mathcal{F}(t')\right> = \Gamma\delta(t - t')\), where \(\Gamma = 2\gamma k_BT\).

The discrete form of Eq.~(\ref{leni}) is
\begin{equation}
R(t + \Delta t) = R(t) -\frac{F_{dna}(r)\Delta t}{\gamma} +
\frac{\sqrt{\Gamma\Delta t}}{\gamma}\xi_n,
\label{le1d}
\end{equation}
where \(\xi_n\) is a Gaussian random variable with \(\left<\xi_n\right>=0\) and \(\left<\xi_n\xi_m\right>=\delta_{nm}\).

Until this point, we have treated the motion of the microsphere using a reference frame directed along the force axis of the DNA molecule. This one-dimensional reference frame is not appropriate to our purposes, since it does not contain  the experimental reference frame which we want to simulate, in which the microsphere undergoes three-dimensional Brownian motion. Therefore, we extend Eq.~(\ref{le1d}) to three-dimensional space, obtaining
\begin{equation}
X_i(t + \Delta t) = X_i(t) -\frac{F_{dna}(r)\Delta t}{\gamma}\frac{X_i(t)}{R} +
\frac{\sqrt{\Gamma\Delta t}}{\gamma}\xi_{n,i},
\label{le3d}
\end{equation}
with \(i = 1,2,3\) and
\begin{equation}
r = \sqrt{\sum_{i=1}^{i=3}X_i^2} - a = R - a,
\label{r}
\end{equation}
where the set \(\{X_i\}\) represents the coordinates of the center of mass of the microsphere while \(r\) represents the relative distance between the two extremities of the DNA molecule. Since the DNA is attached to the surface of the microsphere, the two coordinate systems are connected through Eq.~(\ref{r}), where \(a\) is the radius of the microsphere.

\subsection{Correcting \(\gamma\): The wall-drag effect}

\begin{figure}
\centering
\includegraphics[width=10cm]{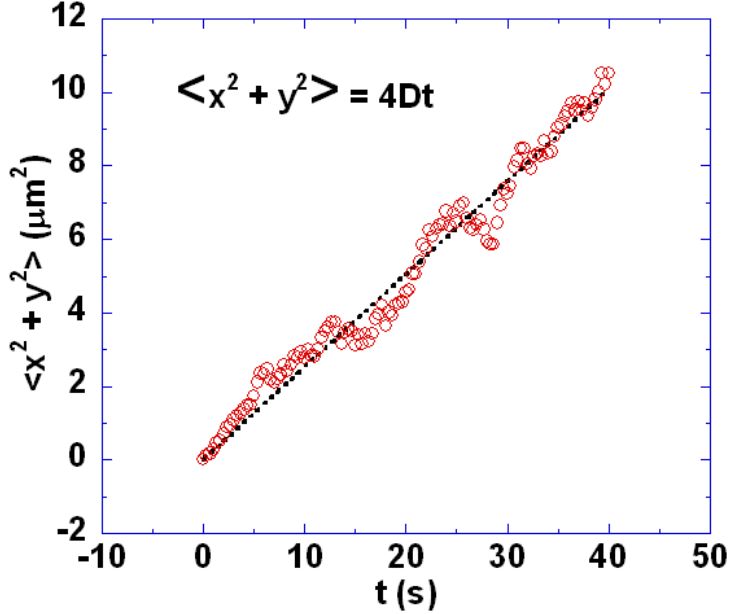}
\caption{Average quadratic displacement as a function of time for microspheres hovering close to the coverglass. The movements of 20 free microspheres were tracked for 40s and combined into a single curve. The difusion coefficient \(D=(0.063 \pm 0.001)\mu m^{2}/s)\) was obtained by fitting this curve to a two-dimensional random walk.}
\label{difusion-xy}
\end{figure}

Use of a constant and isotropic Stokes friction coefficient \(\gamma\) is justified if and only if the fluid velocity attains constant value \(v_0\) far from the microsphere, in all directions. When an obstacle interferes with the fluid flow near the microsphere, such as the coverglass surface in the present case, this assumption is no longer valid and \(\gamma\) depends on the direction of motion and on the distance from the microsphere to the obstacle \cite{Feitosa1}. According to the analyses of Faxen \cite{Faxen1}, Brenner \cite{Brenner1} and Goldman \emph{et al} \cite{Goldman1} in the early and mid 20th century, for a microsphere of radius \(a\) whose center of mass is at a distance \(h\) from a planar surface, the friction coefficient \(\gamma\) is anisotropic, with
\begin{subequations}
\begin{equation}
\gamma_{||} \approx \frac{6\pi\eta a}{\left(1 -
\frac{9}{16}\left(\frac{a}{h}\right) + \frac{1}{8}\left(\frac{a}{h}\right)^3 -
\frac{45}{254}\left(\frac{a}{h}\right)^4 -
\frac{1}{16}\left(\frac{a}{h}\right)^5\right)}
\label{gammah}
\end{equation}
for motion parallel to the planar surface \emph{(\(xy\) plane)}, and
\begin{eqnarray}
\gamma_{\perp} = & \; 6\pi\eta a \; \times & \left[ \frac{4}{3} \sinh\; \alpha \;\sum_{n=1}^{\infty}\frac{n(n+1)}{(2n-1)(2n+3)}\; \times \; \right. \nonumber \\
[-3mm] &  & \label{gammav-e} \\ [-3mm] & &
\left.\left[\frac{2\sinh\left[(2n+1)\alpha\right] + (2n+1)\sinh\;2\alpha}{4\sinh^2\left[(n + \frac{1}{2})\alpha\right] - (2n+1)^2\sinh^2\alpha} - 1\right]\right] \nonumber,
\end{eqnarray}
for motion perpendicular to the planar surface \emph{(z axis)}, where \(\alpha = \cosh^{-1}(h/a)\). The exact correction for \(\gamma_{\perp}\), shown in Eq.~(\ref{gammav-e}), can be replaced by a far simpler approximation, given by \cite{Brenner1}
\begin{equation}
\gamma_{\perp} \approx 6\pi\eta a\left(1 + \frac{a}{h-a}\right),
\label{gammav-a}
\end{equation}
\label{gammas}
\end{subequations}
with an average error \(\leq 5\%\) for \(h/a >1\). The loss of isotropy in \(\gamma\) alters Eq.~(\ref{le3d}), which becomes
\begin{subequations}
\begin{equation}
X_i(t + \Delta t) = X_i(t) -\frac{F_{dna}(r)\Delta
t}{\gamma_{||}}\frac{X_i(t)}{R} + \frac{\sqrt{\Gamma_{||}\Delta
t}}{\gamma_{||}}\xi_{n,i}\; \;\;\; i = 1, 2,
\label{le3d-a}
\end{equation}
\begin{equation}
X_3(t + \Delta t) = X_3(t) -\frac{F_{dna}(r)\Delta
t}{\gamma_{\perp}}\frac{X_3(t)}{R} + \frac{\sqrt{\Gamma_{\perp}\Delta
t}}{\gamma_{\perp}}\xi_n,
\label{le3d-b}
\end{equation}
\label{le3d-c}
\end{subequations}
where Eq.~(\ref{le3d-a}) generates the \emph{x} and \emph{y} components of the recoil curve and Eq.~(\ref{le3d-b}) the \emph{z} component. This equation, however, does not take into account the fact that the height \(h\) of the center of the microsphere cannot be less than its radius \(a\), in which case the microsphere is touching the coverglass. At this point, interactions between the microsphere and the coverglass can no longer be neglected and Eq.~(\ref{gammas}) can no longer be safely used. We have nevertheless observed that, after the recoil, the microsphere exhibits Brownian motion in the plane of the coverglass for quite some time, the same behavior being observed for most of the free microspheres as well. This observation indicates that the microspheres remain hovering close to the coverglass for a period of time before sticking to it. To determine how close to the coverglass the microspheres were hovering, we prepared a sample containing only free microspheres under the same experimental conditions. After letting the sample rest on the microscope for more than 2 hours, we measured the microsphere diffusion coefficient for motion parallel to the coverglass (Fig.~(\ref{difusion-xy})). The measured value was \((0.063 \pm 0.001)\mu m^{2}/s\), approximately \(2.75\) times less than the theoretical free difusion coefficient, \(\frac{k_bT}{6\pi\eta a} = 0.173 \mu m^{2}/s\), for \(a=1.42\mu m\). Assuming this discrepancy is caused by the increased friction close to the coverglass and applying Eq.~(\ref{gammah}) we obtained an average hovering distance of approximately \(1.47\;\mu m\), indicating that the microspheres in fact were not touching the coverglass, their point of nearest approach remaining approximately \(0.05 \mu m\) above it.
The simulation routine was then altered to reproduce this effect, not letting \(h\) drop below its minimum value of \(1.47\mu m\) during the recoil, as observed experimentally.

\begin{figure}
\centering
\includegraphics[width=10cm]{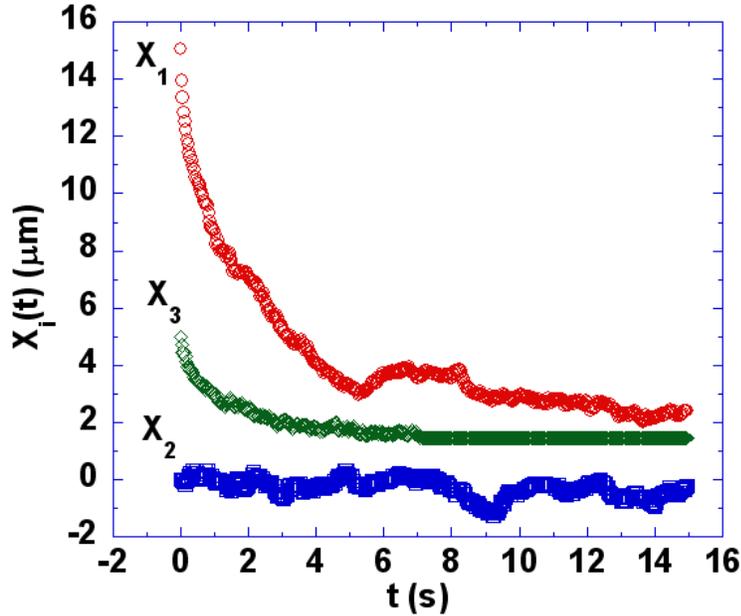}
\caption{A typical simulated recoil curve. The numerical solution of Eq.~(\ref{le3d-c}) is plotted for each direction. For this simulation we used \(h_0 = 5\mu m\), \(A = 45nm\), \(L=17\mu m\) and \(T = 25^oC\).}
\label{simrec}
\end{figure}

The initial conditions of the microsphere for all simulated recoils are \(X_1(0) = \sqrt{\left(0.85L+a\right)^2 - h_0^2}\), \(X_2(0) = 0\) and \(X_3(0) = h_0\), where we have assumed that the DNA molecule is at least \(85\%\) stretched, initially. The coordinate system is taken so as to place the direction of stretching in the \emph{\(xz\) plane}. The time increment, \(\Delta t\), was set to \((1/30)s\), matching videomicroscopy time resolution for better comparison between simulated and experimental curves (which also proved adequate for numerical convergence). A simulated recoil curve is shown in Fig.~(\ref{simrec}).

\section{Experiment}
\label{exp}

\subsection{Experimental Setup}

\begin{figure}
\centering
\includegraphics[width=10cm]{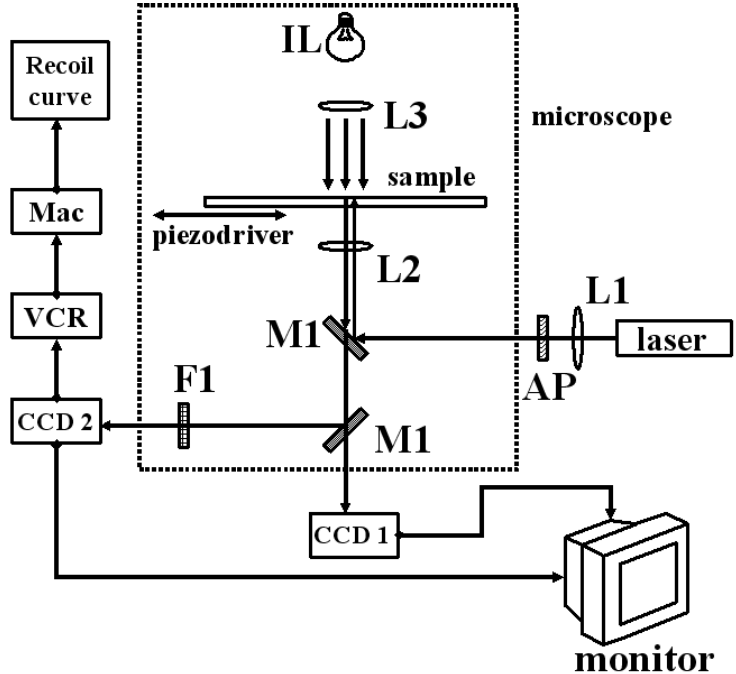}
\caption{Schematic of the experimental setup. \small\textit{(IL) light source, (L1) objective lens $20\times$, (L2) objective lens $100\times$, (L3) condenser, (AP) anamorphic prism, (M1) dicroic mirror, (F1) near infrared filter.}}
\label{setup}
\end{figure}

The experiments are made on a Nikon \emph{TE300 Eclipse} inverted optical microscope coupled to two CCD cameras and a piezo-driven stage (Fig.~\ref{setup}). We use a SDL~5422 near infrared laser \emph{(\(\lambda = 832\:nm\))} focused through an infinity corrected objective lens \emph{(Nikon plan~apo, DIC~H, 100\(\times\), 1.4~NA, oil immersion)} to create the optical tweezers. Imaging from CCD~1 is used to locate and move suitable microspheres into position while imaging from CCD~2, filtered from the near infrared light, is videorecorded for posterior analysis.

\subsection{Sample Preparation}

Samples are prepared using the method described in \cite{Shivashankar2}, with slight variations. We prepare a solution containing PBS \((150\: mM\:Na^+)\) pH 6 and microspheres \emph{(Polybead Polystyrene \(2.8\:\mu m\) microspheres, \(\rho = 1.05\:g/ml, n = 1.6\))}. The density of microspheres in the solution is around \(450/\mu l\). For every \(395\:\mu l\) of solution, we add \(5\:\mu l\) of DNA solution \emph{(New England Biolabs \(\lambda\) DNA \(553\:\mu g/ml\))}, previously heated in a thermal bath at 62\(^oC\) for 5 minutes, mix gently and incubate together for 20 minutes. The solution is then pipeted into cells \emph{(\(\approx\) \(4\:mm\) radius, \(3\:mm\) height)} on coverglasses and covered with cut pieces of coverglass to prevent excessive evaporation. At this pH condition, DNA molecules bind to the microspheres and to the coverglass, preferably by its extremities \cite{Allemand1}. After 24h of incubation at room temperature, the cells are opened and gently washed with PBS \((150\: mM\:Na^+)\) pH 7.4 to remove the excess of free beads and reset the pH for the DNA. The cells are covered again and are then ready for experiments.

\subsection{DNA-Microsphere System Selection}

With the samples placed on the microscope, the optical tweezers are used to locate a microsphere bound to a single DNA molecule. Although many microspheres are attached to DNA, some judgement is needed to choose a good DNA-microsphere system in order to avoid microspheres attached to more than one molecule, molecules bound to the coverglass at multiple locations and other possible undesirable configurations. Verifying if the selected microsphere presents radial symmetry with respect to stretching and recoiling around its apparent resting point is the minimal required test for a good candidate. Once a suitable DNA-microsphere pair is selected, the initial height \(h_0\) for the recoil is set adjusting the focal distance of the tweezers. The microsphere is trapped and moved to stretch the DNA. The videorecording is started and the tweezers cut off, starting the recoil.

\subsection{Data Extraction}

After videorecording the recoil through CCD 2, the images are analyzed using the \emph{Image 1.62} \cite{Image} software package. The \emph{x} and \emph{y} coordinates of the center of mass of the microsphere are extracted from each frame and combined with the time index of the recording to give \(X_1(t)\) and \(X_2(t)\).

\subsection{Estimate for \(h(t)\)}

\begin{figure}
\centering
\includegraphics[width=10cm]{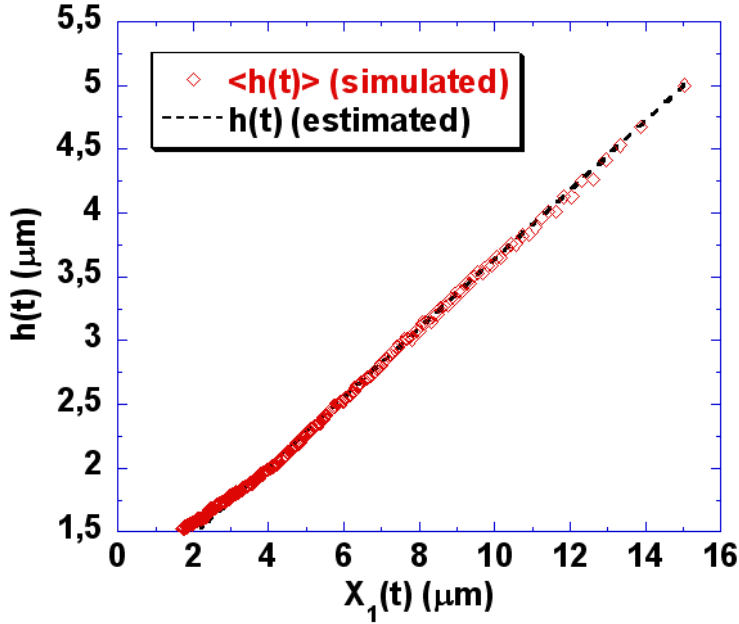}
\caption{The height \(h\) of the center of mass of a microsphere is plotted as a function of its horizontal position. (\(\diamond\)) Average of 250 simulated recoils showing \(\left<h(t)\right>\) as a function of the corresponding \(\left<X_1(t)\right>\). (\(- -\)) Estimate for \(h(t)\), given by Eq.~(\ref{h-eq}). We have tested  Eq.~(\ref{h-eq}) with different values of \(h_0\), \(A\), \(L\) and temperature and obtained similar results. Parameters: \(h_0 = 5\mu m\), \(A = 45nm\), \(L=17\mu m\) and \(T = 25^oC\).}
\label{h-graph}
\end{figure}

As already mentioned, the appropriate correction to the friction coefficient for the microsphere depends on the distance \(h\) of its center from the coverglass. We use the optical tweezers to set the initial height \(h_0\) of the microsphere. During the recoil, the value of \(h\) can be measured from the defocusing of the microsphere, a very time consuming procedure. We believe that, for the recoil experiments, sufficient accuracy can be achieved by using a reasonable estimate for \(h(t)\) guided by computer simulations. To obtain \(h(t)\) for the experimental recoils we used averaged simulated recoils to establish how \(h(t)\), which corresponds to \(X_3(t)\) in the simulated recoils, was related to the horizontal component of the recoil, \(X_1(t)\). We found that, because of the anisotropy in \(\gamma\), \(h(t)\) plotted as a function of \(X_1(t)\) does not correspond to a straight line trajectory pointing to the origin of the coordinate system, as would be expected if \(\gamma_{||}\) and \(\gamma_{\perp}\) were identical. Nevertheless, we observed that the average \(h(t)\) can be estimated, with good agreement  with the simulations (Fig.~(\ref{h-graph})), by
\begin{equation}
h(t) = 0.925 + (h_0 - 0.925)\frac{X_1(t)}{X_1(t=0)},
\label{h-eq}
\end{equation}
until \(h(t)\) reaches \(1.47\mu m\), where the experimental evidence indicates that the microspheres interrupt, at least for a time longer than the duration of the recoil, their vertical motion. During the most relevant region of the recoil, the distance from the bottom of the microsphere to the coverglass is larger than \(0.5\mu m\), a value much larger than any roughness of such a good optical quality coverglass used.

\section{Data Analysis}
\label{analysis}

Using the previously chosen coordinate system, in which the molecule is stretched along the \emph{\(xz\) plane}, the recoil is primarily along the \emph{\(x\) axis}. We may then use the \(x\) component of Eq.~(\ref{eqm}), given by
\begin{equation}
\frac{dX_{1}}{dt} =
-\frac{F_{dna}(r)}{\gamma_{//}}\frac{X_{1}}{R},
\label{fit}
\end{equation}
to fit our data. The \(y\) component of the recoil will fluctuate around 0, allowing us to write
\begin{equation}
R \approx \sqrt{X_{1}^2 + h^2}.
\label{R}
\end{equation}

\begin{figure*}
\centering
\includegraphics[width=15cm]{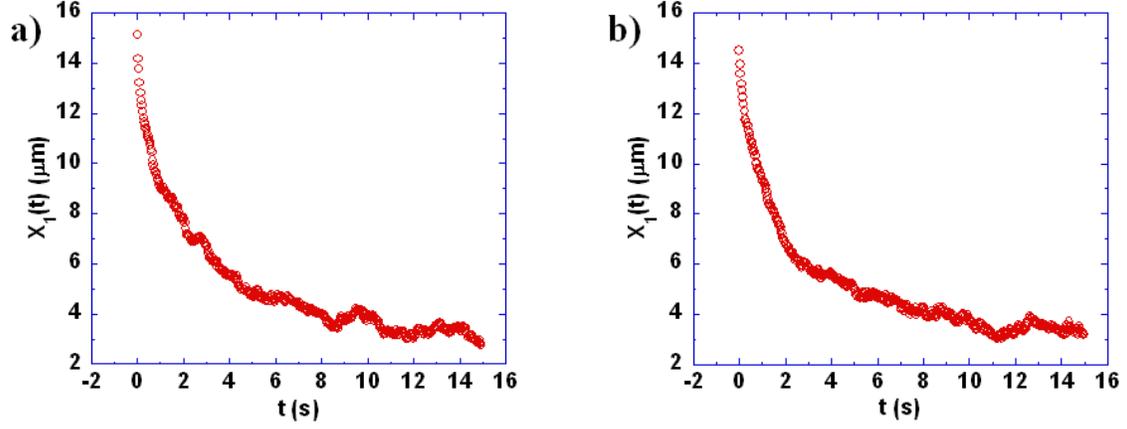}
\caption{Simulated (a) and experimental (b) recoil curves. The parameters of the simulated curve were chosen close to the values obtained from the analysis of the experimental recoil, in order to show that the simulated recoils are a good reproduction of the experiments. The following values were used: \(A=45nm\), \(L=16.5\mu m\), \(h_0=3\mu m\) and \(T=23^oC\).}
\label{recoils}
\end{figure*}

Analysis of the recoil curves (Fig.~(\ref{recoils})) consists of three steps. First, the recoil curve is \emph{smoothed} using the \emph{Stineman} smoothing function from \emph{KaleidaGraph} \cite{Kaleida} software package, filtering out high-frequency Brownian noise. The smoothed recoil curve is then used to evaluate the finite difference \(\frac{\Delta X_1}{\Delta t}\). The resulting velocity data are fit according to Eq.~(\ref{fit}).

\begin{figure*}
\centering
\includegraphics[width=15cm]{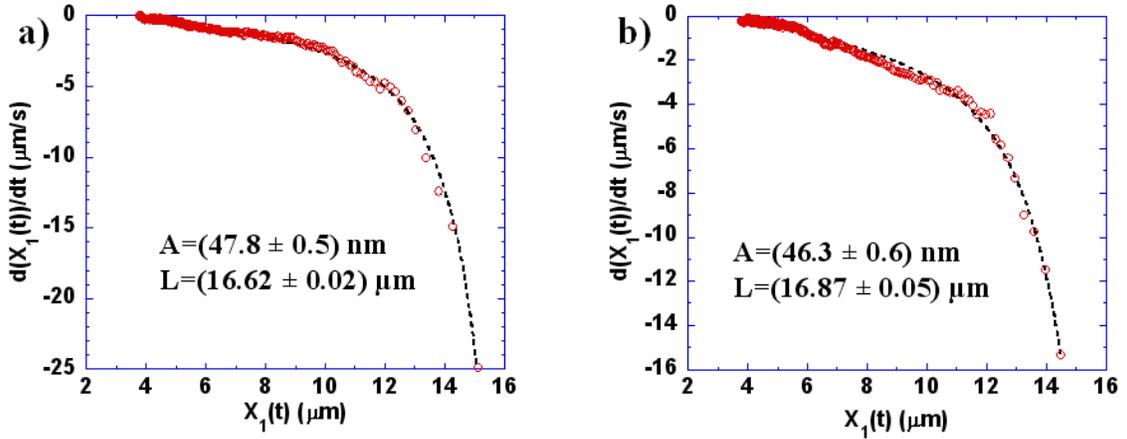}
\caption{Analysis of the recoils shown in Fig.~(\ref{recoils}). a) Simulated recoil. b) Experimental recoil. For the simulated recoils, differences between the parameters \(A\) and \(L\) used in the simulation and the values obtained from the analysis of the recoil are considered acceptable if \(\leq 5nm\) for \(A\) and \(\leq 0.3\mu m\) for \(L\).}
\label{fits}
\end{figure*}

To check the validity of the results produced by this procedure, we compared the values of \(A\) and \(L\) obtained from the analysis of several simulated recoils with the known values of \(A\) and \(L\) used as parameters in each simulation. The process was repeated several times, with different values of \(A\), \(L\), \(h_0\) and \(T\) used in the simulated recoils. In most cases, there was good agreement between the values of \(A\) and \(L\) used in the simulation and the values obtained from the analysis of the recoils generated (Fig.~(\ref{fits})). We have, however, observed that in certain cases the analysis of a simulated recoil produced results quite different than those obtained from the analysis of other recoils generated with the same parameters. We believe that such occasional discrepancies occur when a slight deviation from the average recoil trajectory is enhanced when the recoil is smoothed and its derivative evaluated. To minimize this effect in the experimental analysis, we have always recorded, analysed and compared at least 4 recoils for each DNA-microsphere complex.

We analysed 21 experimental recoils from 5 different DNA-microsphere systems in 3 different samples. The average persistence length obtained from our data was \(A = (43 \pm 5)nm\), a value well within the range obtained from other techniques \cite{Wang1, Viana1, Smith1}.

\section{Conclusion}
\label{conclusion}

In recent years measurements of the persistence length of single DNA molecules have been made through several different methods, ranging from spectroscopic techniques to electron microscopy. Each of these approaches involves a particular set of assumptions, as pointed out by Wang \emph{et al.} in \cite{Wang1}, and is susceptible to different sources of error.

We have shown, based on simulated and experimental data, that the \emph{recoil method}, in which we follow the recoil dynamics of a previously stretched DNA molecule through the motion of a tethered microsphere, can be used to determine the persistence length of single DNA molecules, provided that adequate correction of the friction for the microsphere is made. Although it is possible to perform the analysis of the recoil using only on the horizontal component of the movement, knowledge of the vertical dynamics of the microsphere is fundamental, since its vertical position is the critical parameter governing the friction correction. Both simulated and experimental data were treated using the quasistatic response of the DNA molecule, without considering the far from equilibrium dynamics proposed in \cite{Bohbot-Raviv1}. We observe a high degree of consistency between experiment and simulation.

\begin{ack}

We acknowledge helpful discussions with G. V. Shivashankar and M. Feingold. This work was supported by the Brazilian Agencies: Funda\c c\~ao de Amparo \`a Pesquisa do Estado de Minas Gerais (FAPEMIG), Conselho Nacional de Desenvolvimento Cient\'\i fico e Tecnol\'ogico (CNPq) and FINEP-PRONEX.

\end{ack}
\bibliographystyle{elsart-num}
\bibliography{dna-wlc}

\begin{thebibliography}{10}
\expandafter\ifx\csname url\endcsname\relax
  \def\url#1{\texttt{#1}}\fi
\expandafter\ifx\csname urlprefix\endcsname\relax\def\urlprefix{URL }\fi

\bibitem{Strick1}
T.~Strick, J.-F. Allemand, V.~Croquette, D.~Bensimon, Prog. Biophys. Mol. Bio.
  74 (2000) 115.

\bibitem{Bustamante1}
C.~Bustamante, J.~F. Marko, E.~D. Siggia, Science 265 (1994) 1599.

\bibitem{Marko1}
J.~F. Marko, E.~D. Siggia, Macromolecules 28 (1995) 8759.

\bibitem{Bouchiat1}
C.~Bouchiat, M.~D. Wang, J.-F. Allemand, T.~Strick, S.~M. Block, V.~Croquette,
  Biophys. J. 76 (1999) 409.

\bibitem{Ashkin1}
A.~Ashkin, Proc. Natl. Acad. Sci. USA 94 (1997) 4853.

\bibitem{Wang1}
M.~D. Wang, H.~Yin, R.~Landick, J.~Gelles, S.~M. Block, Biophys. J. 72 (1997)
  1335.

\bibitem{Shivashankar1}
G.~V. Shivashankar, G.~Stolovitzky, A.~Libchaber, Appl. Phys. Lett. 73 (1998)
  291.

\bibitem{Shivashankar2}
G.~V. Shivashankar, M.~Feingold, O.~Krichevsky, A.~Libchaber, Proc. Natl. Acad.
  Sci. USA 96 (1999) 7916.

\bibitem{Viana1}
N.~B. Viana, R.~T.~S. Freire, O.~N. Mesquita, Phys. Rev. E 65 (2002) 041921.

\bibitem{Feingold1}
M.~Feingold, Physica E 9 (2001) 616.

\bibitem{Bohbot-Raviv1}
Y.~Bohbot-Raviv, W.~Zhao, M.~Feingold, C.~H. Wiggins, R.~Granek, Phys. Rev.
  Lett. 92 (2004) 098101.

\bibitem{Perkins1}
T.~T. Perkins, S.~R. Quake, D.~E. Smith, S.~Chu, Science 264 (1994) 822.

\bibitem{Feitosa1}
M.~I.~M. Feitosa, O.~N. Mesquita, Phys. Rev. A 44 (1991) 6677.

\bibitem{Kampen1}
N.~G.~v. Kampen, Stochastic Processes in Physics and Chemistry, North-Holland
  Publishing Company, 1981.

\bibitem{Doi1}
M.~Doi, S.~F. Edwards, The Theory of Polymer Dynamics, Oxford University Press,
  1986.

\bibitem{Faxen1}
H.~Faxen, Ark. Mat. Astron. Fys. 18 (1924) 1.

\bibitem{Brenner1}
H.~Brenner, Chem. Eng. Sci. 16 (1961) 242.

\bibitem{Goldman1}
A.~J. Goldman, R.~G. Cox, H.~Brenner, Chem. Eng. Sci. 22 (1967) 637.

\bibitem{Allemand1}
J.-F. Allemand, D.~Bensimon, L.~Jullien, A.~Bensimon, V.~Croquette, Biophys. J.
  73 (1997) 2064.

\bibitem{Image}
NIH Image 1.62 \textit{http://rsb.info.nih.gov/nih-image}.

\bibitem{Kaleida}
KaleidaGraph 3.5 \textit{http://www.kaleidagraph.com}.

\bibitem{Smith1}
S.~B. Smith, L.~Finzi, C.~Bustamante, Science 258 (1992) 1122.

\end{thebibliography}

\end{document}